\documentclass[useAMS,usenatbib]{mn2e}

\usepackage{float} % If comment this, figure moves to Page 2
\usepackage{epsfig}
\usepackage{graphicx}
\usepackage[latin1]{inputenc}
\usepackage{latexsym}
\def\simlt{\ \raise -2.truept\hbox{\rlap{\hbox{$\sim$}}\raise5.truept   %
\hbox{$<$}\ }}
\def\simgt{\ \raise -2.truept\hbox{\rlap{\hbox{$\sim$}}\raise5.truept   %
\hbox{$>$}\ }}                                                          %

\def\be{\begin{equation}}
\def\ee{\end{equation}}
\def\newline{\hfil\break}

\def\la{\mathrel{\hbox{\rlap{\hbox{\lower4pt\hbox{$\sim$}}}\hbox{$<$}}}}
\def\ga{\mathrel{\hbox{\rlap{\hbox{\lower4pt\hbox{$\sim$}}}\hbox{$>$}}}}

\def\mach{\mathcal{M}}
\newcommand{\pd}[3]{\frac{\partial^{#3} #1}{\partial {#2}^{#3}}} %partial derivative of #1 w.r.t #2 of order #3
\newcommand{\td}[3]{\frac{d^{#3} #1}{d {#2}^{#3}}} %total derivative of #1 w.r.t #2 of order #3
\renewcommand{\v}[1]{\ensuremath{\mathbf{#1}}} % for vectors
\newcommand{\gv}[1]{\ensuremath{\mbox{\boldmath $ #1 $}}}
\title[Dark Matter in A520]{The role of Dark Matter annihilation in the radio emission of the galaxy cluster A520}
\author[Marchegiani, Colafrancesco and Khanye]{P. Marchegiani$^{1}$\thanks{E-mail: Paolo.Marchegiani@wits.ac.za}, S. Colafrancesco$^{1}$\thanks{Deceased} and N.F. Khanye$^{1}$\\
$^{1}$School of Physics, University of the Witwatersrand, Private Bag 3, 2050-Johannesburg, South Africa\\
}

\begin{document}

\date{Accepted: 2018 November 30; in original form: 2018 October 10}

\pagerange{\pageref{firstpage}--\pageref{lastpage}} \pubyear{2018}

\maketitle

\label{firstpage}

\begin{abstract}
A520 is a hot and luminous galaxy cluster, where gravitational lensing and X-ray measures reveal a different spatial distribution of baryonic and Dark Matter. This cluster hosts a radio halo, whose map shows a separation between the North-East and the South-West part of the cluster, similarly to what is observed in gravitational lensing maps. In this paper we study the possibility that the diffuse radio emission in this cluster is produced by Dark Matter annihilation. We find that in the whole cluster the radio emission should be dominated by baryonic phenomena; if a contribution from Dark Matter is present, it should be searched in a region in the NE part of the cluster, where a peak of the radio emission is located close to a Dark Matter sub-halo, in a region where the X-ray emission is not very strong. By estimating the radio spectrum integrated in this region using data from publicly available surveys, we find that this spectrum can be reproduced by a Dark Matter model for a neutralino with mass 43 GeV and annihilation final state $b \bar b$ for a magnetic field of 5 $\mu$G.
\end{abstract}

\begin{keywords}
Cosmology: Dark Matter; Galaxies: clusters: general, theory.
\end{keywords}

%%%%%%%%%%%%%%%%%%%%%%%%%%%%%%%%%%%%%%%%%%%%%%%

\section{Introduction}     

Galaxy clusters are supposed to be dominated by a component of Dark Matter (DM) of unknown nature. The annihilation of DM particles in a galaxy cluster can produce relativistic electrons and gamma rays, whose possible detection can be used to test the properties of the DM particles and the cluster itself (Colafrancesco, Profumo \& Ullio 2006). However, other processes in galaxy clusters can produce diffuse non-thermal emission in several spectral bands, like radio, X-rays, and gamma rays (e.g. Feretti et al. 2012), making difficult to distinguish the emission of baryonic origin from the one originated by DM annihilation. Marchegiani \& Colafrancesco (2015) proposed to address this problem by considering clusters where the X-rays and gravitational lensing measures suggest that the baryonic and the DM halos are located in different positions, like in the case of the Bullet cluster (Clowe et al. 2006), and studying the emission detected in the different regions of the cluster. This situation can usually be found in clusters during a merging event taking place approximately perpendicularly to the line of sight, when the DM halos of the two clusters are expected to cross each other in a collisionless way, whereas the thermal gas halos interact by ram pressure, producing a slowdown and an offset between DM and baryonic matter (e.g. Roettiger, Locken \& Burns 1997).  

A520 is a hot and luminous galaxy cluster, where an ongoing merging event has been identified both with X-rays (Markevitch et al. 2005) and optical observations (Proust et al. 2000). 
In A520 X-rays and gravitational lensing measures reveal a situation similar to the case of the Bullet cluster: the distribution of the DM density in fact results to be different from the baryonic one, with two main DM peaks located on the opposite sides of the cluster geometrical center along the NE-SW direction, that is interpreted as the merging axis, whereas the X-ray emission shows a main peak close to the cluster geometrical center, and an extension in the SW part of the cluster, where a second peak, associated to the Brightest Cluster Galaxy (BCG) of the secondary sub-cluster, is located close to the edge of the cluster, and presents a trail of cold gas, interpreted as the remnant of a cool core cluster that is going to be destroyed by ram pressure while crossing the main cluster (Clowe et al. 2012; Wang, Markevitch \& Giacintucci 2016; Wang, Giacintucci \& Markevitch 2018). The detection of a third peak in the DM distribution close to the X-ray peak is controversial: in some studies (e.g. Mahdavi et al. 2007) this peak results to be present, whereas in other studies (e.g. Clowe et al. 2012) this peak is not detected with sufficient statistical significance.

A520 hosts a giant bright radio halo (Govoni et al. 2001; Govoni et al. 2004; Vacca et al. 2014; Wang et al. 2018), with a relatively flat spectral index (with integrated value between 325 and 1400 MHz of $\alpha\sim1.12$; Vacca et al. 2014) that, although with fluctuations in its value across the cluster, does not present a systematic steepening towards the peripheral regions (Vacca et al. 2014), as instead observed in other clusters like Coma (Giovannini et al. 1993) and A665 (Feretti et al. 2004). The radio maps show that the halo is divided in two main sub-halos in the NE and SW regions of the cluster; this structure is interesting because it is analogue to the structure of the DM distribution in the cluster. We note that a similarity between radio halo and DM morphology is also observed, for example, in the Coma cluster (Brown \& Rudnick 2011; Marchegiani \& Colafrancesco 2016).

In this paper we check the possibility that the radio emission in A520 can be originated by DM annihilation. In Sect.2 we study the emission expected from DM in the whole cluster, and compare it with the emission of baryonic origin. In Sect.3 we focus on a region of the cluster where a peak of the diffuse radio emission is located close to a DM peak, and in Sect.4 we study the high-energy emission expected from the cluster. In Sect.5 we discuss our results and make hypothesis on the origin of the electrons in the different regions of the cluster, and in Sect.6 we summarize our conclusions.

Throughout the paper, we use a flat, vacuum--dominated cosmological model following the results of Planck, with $\Omega_m = 0.308$, $\Omega_{\Lambda} = 0.692$ and $H_0 =67.8$ km s$^{-1}$ Mpc$^{-1}$ (Ade et al. 2016). With these values the luminosity distance of A520 at $z=0.199$ is  $D_L=1005$ Mpc, and 1 arcmin corresponds to 203 kpc at this distance.

\section{The integrated emission from the whole cluster}

As a first step, we study the radio emission expected in the whole cluster for a baryonic model, where secondary electrons are produced by hadronic interactions between cosmic rays and thermal protons (e.g. Blasi \& Colafrancesco 1999), and a DM model, where electrons are produced by DM annihilation (e.g. Colafrancesco et al. 2006).

In both the models, the equilibrium spectrum of electrons is obtained by solving the diffusion-loss equation (e.g. Sarazin 1999):
\begin{eqnarray}
\pd{}{t}{}\td{n_e}{E}{} & = & \; \gv{\nabla} \left( D(E,\v{r})\gv{\nabla}\td{n_e}{E}{}\right) + \pd{}{E}{}\left( b(E,\v{r}) \td{n_e}{E}{}\right) + \nonumber \\
& & + Q_e(E,\v{r}) \; ,
\end{eqnarray}
where $(dn_e)/(dE)$ is the electron spectrum, $D(E,\v{r})$ is the spatial diffusion coefficient, $b(E,\v{r})$ is the energy-loss function, and $Q_e(E,\v{r})$ is the electron source function, that can be calculated according to the physics of the hadronic interactions (e.g. Moskalenko \& Strong 1998) or the DM annihilation (e.g. Gondolo et al. 2004).

In the hadronic model, the source term is proportional to the product of the numerical density of thermal nuclei and the one of cosmic ray protons.
For the density of the thermal gas we use the properties derived from X-rays observations: the radial profile can be described by a beta model
\begin{equation}
n_{th}(r)=n_{th,0} \left[1+\left(\frac{r}{r_c}\right)^2\right]^{-q_{th}}
\end{equation}
(Cavaliere \& Fusco-Femiano 1976) with $n_{th,0}=3.8\times10^{-3}$ cm$^{-3}$, $r_c=413$ kpc, and $q_{th}=1.3$ (see table 6 in Govoni et al. 2001, after correcting for the different cosmological model). We assume that the cosmic ray protons have a spatial profile proportional to the thermal one, use power-law spectral distributions $N_p\propto \gamma^{-s_p}$ with several values of the spectral index, and let their central density as a free parameter, expressed through the ratio between the non-thermal and the thermal pressure (see, e.g., Marchegiani, Perola \& Colafrancesco 2007). This last value has been constrained to have upper limits of the order of 2-6\%, depending on the spectral index, from stacked analysis of Fermi gamma ray upper limits in galaxy clusters (Huber et al. 2013; Prokhorov \& Churazov 2014). At the moment there are not estimates of the magnetic field in A520 derived from Farady Rotation Measures available in literature; we therefore adopt a magnetic field similar to the one found in Coma, with a central value of 5 $\mu$G and a radial profile proportional to $n_{th}(r)^{1/2}$ (Bonafede et al. 2010).

For the DM models, we use two neutralino models with mass 9 GeV and annihilation final state $\tau^+\tau^-$, and mass 43 GeV and annihilation final state $b \bar b$, with the total normalizations, given by the product of the annihilation cross section and the substructures boosting factor, ${\cal B} \times \langle \sigma v \rangle$, as obtained from the fitting to the flux of the radio halo in the Coma cluster (Marchegiani \& Colafrancesco 2016). These normalizations can therefore be considered as upper limits on the DM emission, because they can be obtained for optimistic values of the annihilation cross section as derived from the gamma ray excess in the Galactic Center (Abazajian \& Keeley 2016) and the substructures boosting factor (see discussion in Marchegiani \& Colafrancesco 2016), and because a higher value of this normalization would produce a radio emission in excess with respect to the flux observed in the Coma cluster. The density and the radial profile of the DM halo are assumed to have a Navarro, Frenk \& White (1996) profile,
\begin{equation}
\rho(r)=\frac{\rho_s}{\left(\frac{r}{r_s}\right)\left(1+\frac{r}{r_s}\right)^2},
\label{dens.dm}
\end{equation}
with the parameters derived from the total mass, $M_{200}=(9.1\pm1.9)\times10^{14} M_\odot$ (Clowe et al. 2012), according to the procedure described in Bullock et al. (2001) and Colafrancesco, Marchegiani \& Beck (2015): $r_s=472$ kpc and $\rho_s=5.29\times10^3\rho_c$, where $\rho_c$ is the critical density of the Universe. We calculate the total flux by integrating in a volume with a radius of 0.5 Mpc, similar to the radius of the whole radio halo.

\begin{figure}
\centering
\begin{tabular}{c}
\includegraphics[width=\columnwidth]{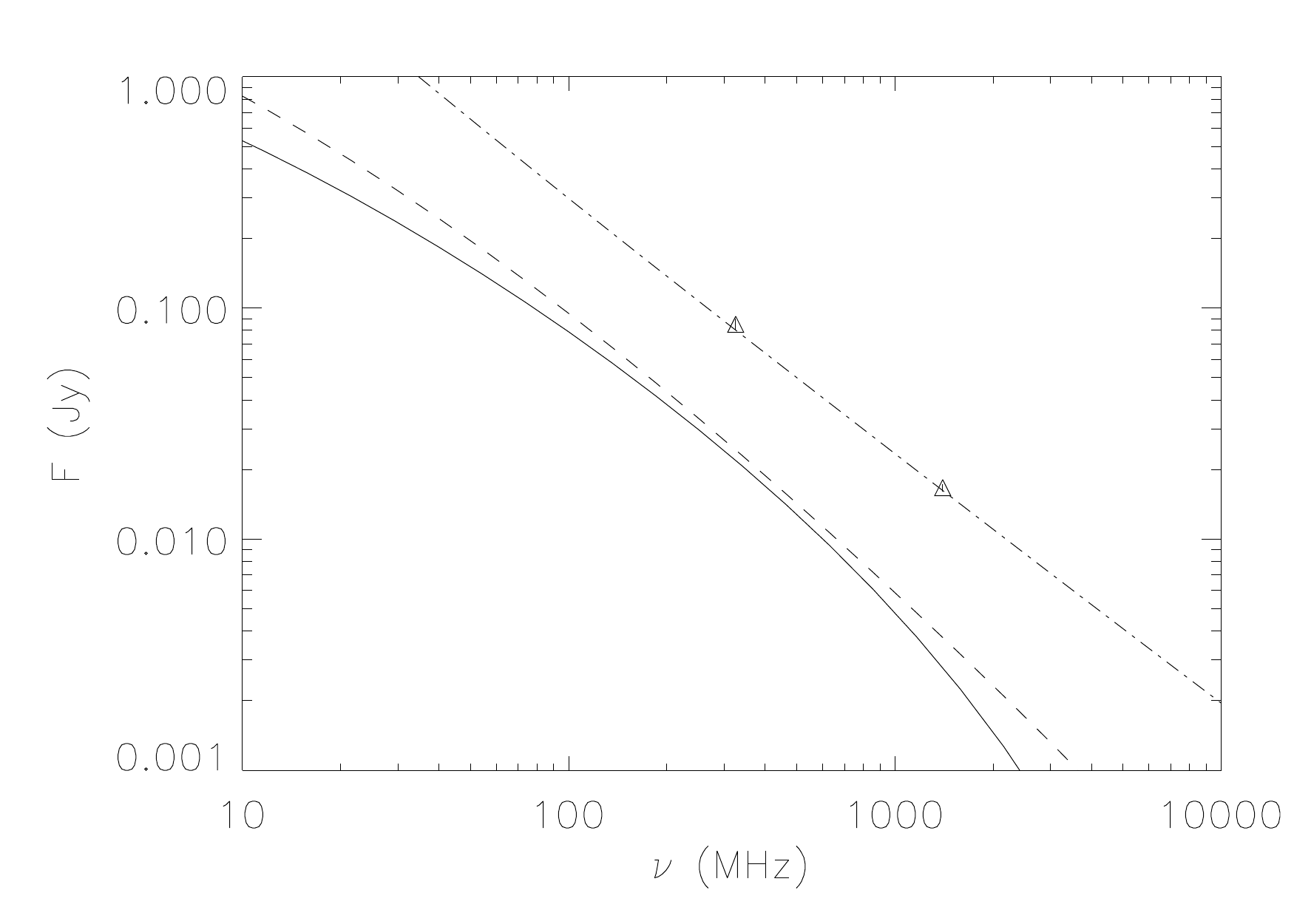}
\end{tabular}
\caption{Spectrum of the radio halo flux in A520 for: \textit{i)} a neutralino model with mass $M_\chi=9$ GeV, annihilation final state $\tau^+\tau^-$, ${\cal B} \times \langle \sigma v \rangle=6\times10^{-25}$ cm$^3$ s$^{-1}$ (solid line); \textit{ii)} a neutralino model with mass $M_\chi=43$ GeV, annihilation final state $b \bar b$, ${\cal B} \times \langle \sigma v \rangle=4\times10^{-24}$ cm$^3$ s$^{-1}$ (dashed line); \textit{iii)} a hadronic model with proton spectral index $s_p=2.2$ and a pressure ratio $P_{CR}/P_{th}=2.6\%$ (dot-dashed line). Data are from Vacca et al. (2014).}
\label{halo.dm+sec}
\end{figure}

The comparison between the prediction of these models and the observed total flux of the radio halo in A520 at 325 and 1400 MHz (Vacca et al. 2014) is given in Fig.\ref{halo.dm+sec}. The hadronic model provides a good fit to the data for a spectral index $s_p=2.2$ and a pressure ratio $P_{CR}/P_{th}=2.6\%$, that is of the order of the upper limits derived from stacked analysis of gamma ray emission in clusters. The DM emission is instead lower than the observed flux by almost one order of magnitude, suggesting that the bulk of radio emission in A520 can not be of DM origin. It is anyway possible that the DM can give a contribution in some regions of the cluster, as suggested in the case of the Bullet cluster (Marchegiani \& Colafrancesco 2015). In the next Section we explore this possibility.

\section{Emission in the DM peak region}

According to Clowe et al. (2012), the two most massive DM subhalos inside the radio halo region in A520 are the ones labelled with P2 and P4, that are located respectively in the NE and SW parts of the halo. The subhalo P4 is located close to the BCG that is moving outwards producing the detected bow shock and leaving a cold gas trail; it is therefore reasonable to think that the radio emission in this region is dominated by the electrons (primary or secondary) originated by this galaxy or accelerated by the associated bow shock.

\begin{figure}
\centering
\begin{tabular}{c}
%\vbox{
\includegraphics[width=\columnwidth]{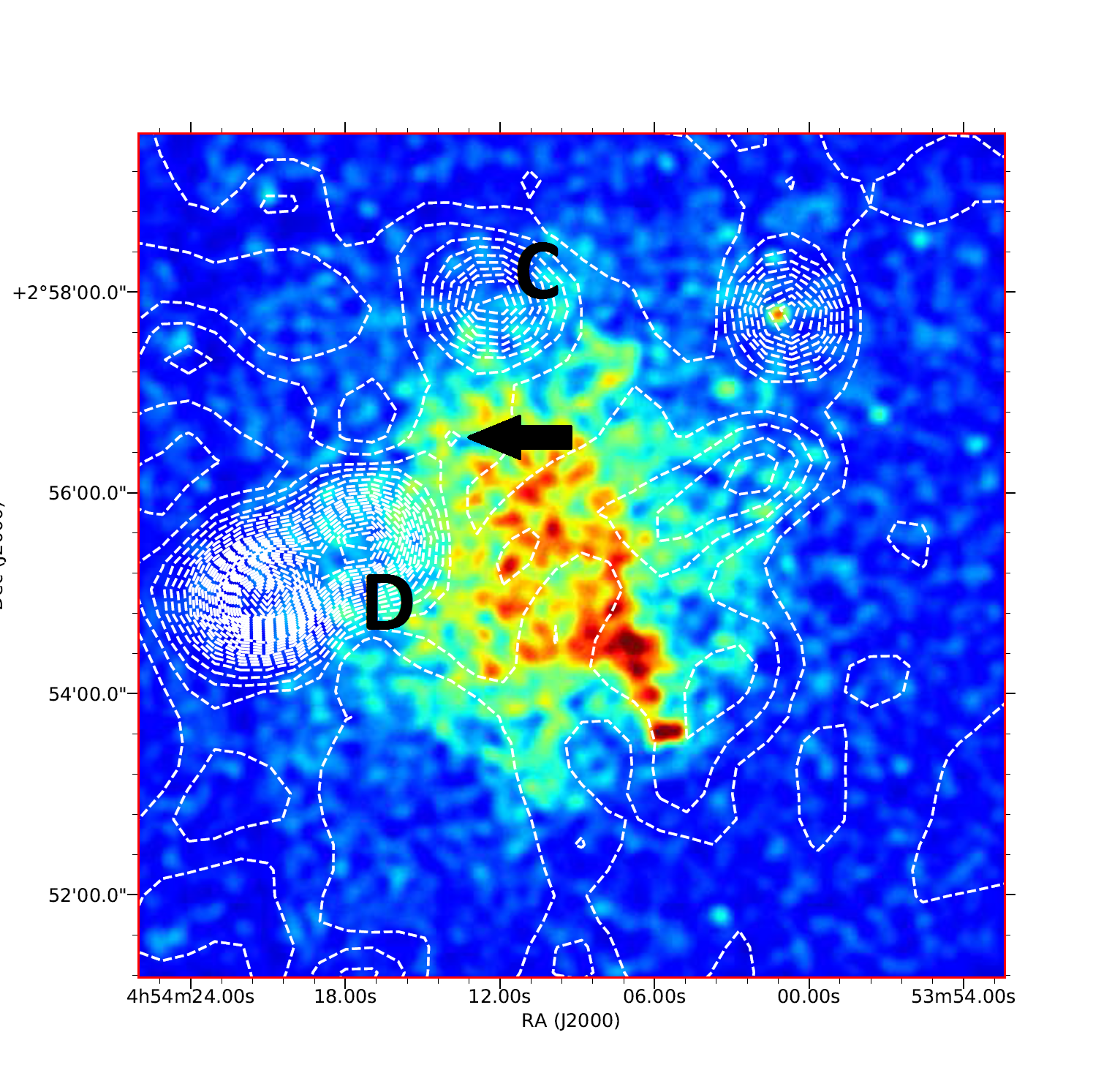}
\end{tabular}
\caption{X-ray image from Chandra with NVSS contours overlapped (white); the black arrow indicates the peak of the NVSS map close to the P2 region; labels C and D are for radio galaxies as in Vacca et al. (2014).
}
\label{map_Xray_radio}
\end{figure}

\begin{figure}
\centering
\begin{tabular}{c}
\includegraphics[width=\columnwidth]{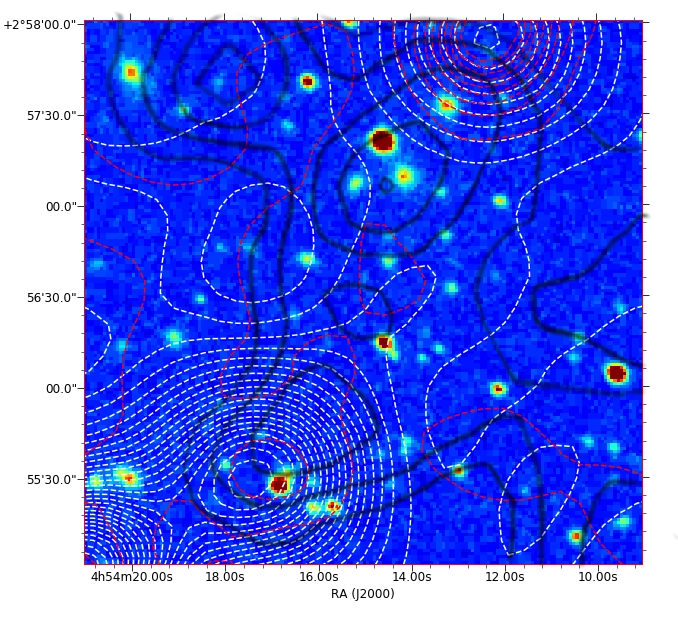}
\end{tabular}
\caption{Optical image from ESO DSS with NVSS (white), TGSS (red), and from weak lensing (from Wang et al. 2016; black) overlapped.}
\label{map_optical_radio}
\end{figure}

The subhalo P2 is instead located in a region with a weak X-ray emission, but close to a peak of the radio emission, located between the radio galaxies labelled with C and D by Vacca et al. (2014) and visible also in the NVSS map (see Fig.\ref{map_Xray_radio}). This radio peak is better visible in the fig. 2a in Wang et al. (2018), where the diffuse radio emission is shown after the removal of point sources, at the approximated coordinates (J2000) RA=04 54 15.0, Dec=+02 56 31.4. In Fig.\ref{map_optical_radio} we show a zoom of this region, with the optical ESO-DSS map, where the contours from NVSS (Condon et al. 1998) and TGSS (Intema et al. 2017) radio surveys, and from the gravitational lensing analysis (from fig. 1c in Wang et al. 2016) are overlapped. The radio peak located close to the center of the image is visible in the TGSS contours, and is located between two lensing peaks; the distance between the center of the TGSS peak and the lensing peaks is of the order of the TGSS resolution (25 arcsec), and therefore it is possible that in this peak there is a contribution from DM annihilation. At the center of the TGSS peak there is a galaxy, presently not recognized as a radio source, that can also be the origin of the radio emission; this situation is similar to the one in the DM Eastern peak in the Bullet cluster, where the radio peak is located close to the DM peak but also to a possible radio galaxy (see Marchegiani \& Colafrancesco 2015).

In order to estimate the radio flux coming from this region, we analyzed the data from a few radio surveys, i.e. VLSSr (Lane et al. 2014), TGSS, and NVSS, estimating the diffuse flux coming from a circle centered at the coordinates RA=04 54 15.0, Dec=+02 56 31.4, and a radius of 35 arcsec (i.e. avoiding the emission from radio galaxies C and D), corresponding to 118 kpc. The data analysis has been done using the CASA package, and the results are reported in Table \ref{p2fluxes}.

\begin{table}
\centering
\begin{tabular}{ccc} 
\hline
$S_{74}$ (mJy) & $S_{150}$ (mJy) & $S_{1400}$ (mJy)\\
\hline
$14.9\pm5.2$ & $7.01\pm0.13$ & $0.450\pm0.047$\\
\hline
\end{tabular}
\caption{Radio fluxes estimated in a circular region centered in RA=04 54 15.0, Dec=+02 56 31.4, and with a radius of 35 arcsec, as obtained from the surveys VLSSr (74 MHz), TGSS (150 MHz), and NVSS (1.4 GHz).} 
\label{p2fluxes}
\end{table}

Then we calculated the hadronic and the DM emission in this region by using the models described in the previous Section. According to the magnetic field model used in the previous calculations, we should expect at the distance of the P2 region from the cluster center a magnetic field of the order of $\sim3.8$ $\mu$G and a thermal gas density of $\sim1.6\times10^{-3}$ cm$^{-3}$. Since the size of this region is small compared to the core radius of the cluster, we use a constant magnetic field and constant densities of thermal electrons and non-thermal protons inside the region.
According to Clowe et al. (2012), the mass of the P2 sub-halo is $M=4.08\times10^{13}$ M$_\odot$, from which the parameters for a NFW model $r_s=112$ kpc (similar to the radius of the region we analyzed) and $\rho_s=1.37\times10^4 \rho_c$ can be derived.

\begin{figure}
\centering
\begin{tabular}{c}
\vbox{
\includegraphics[width=\columnwidth]{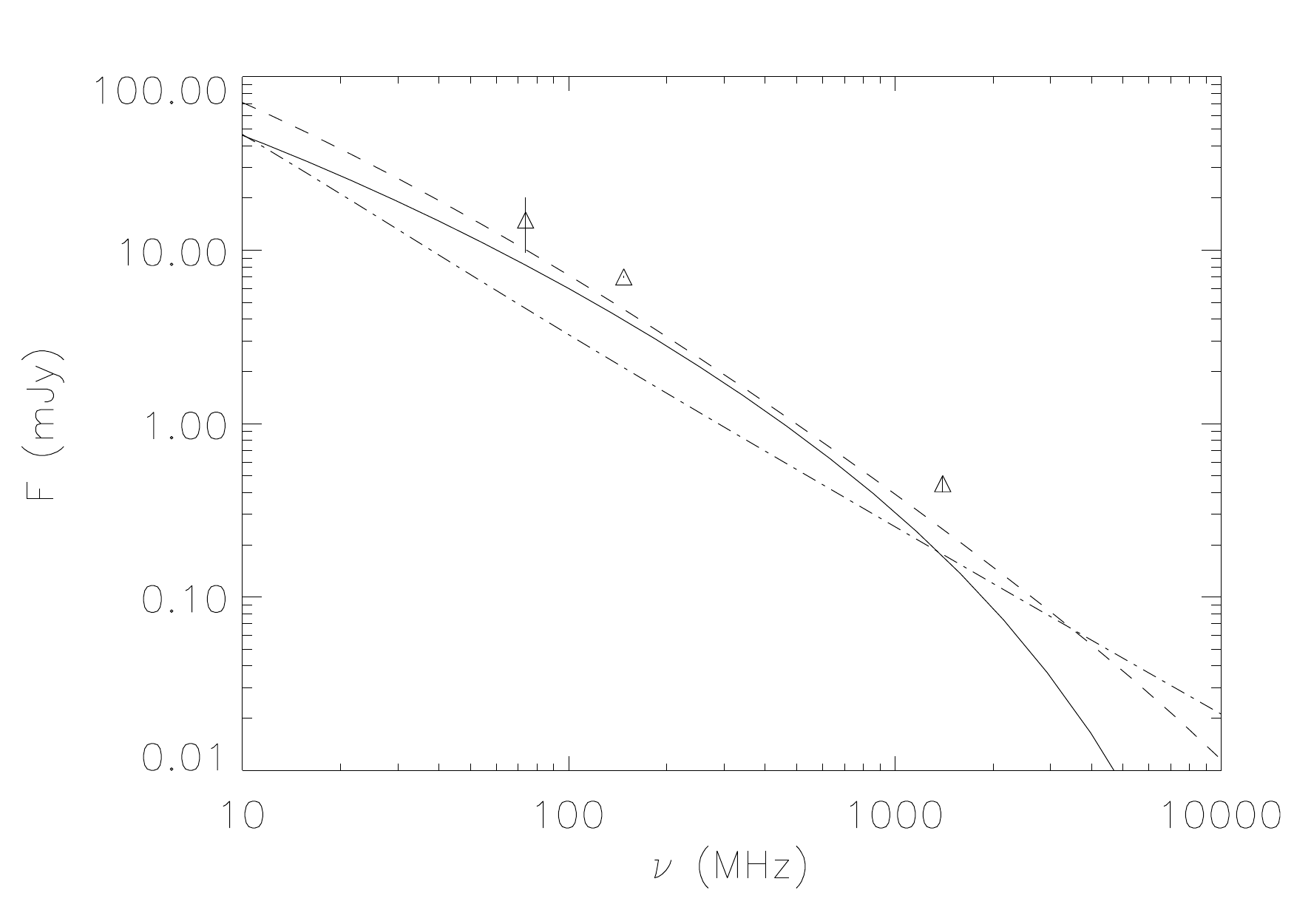}
\includegraphics[width=\columnwidth]{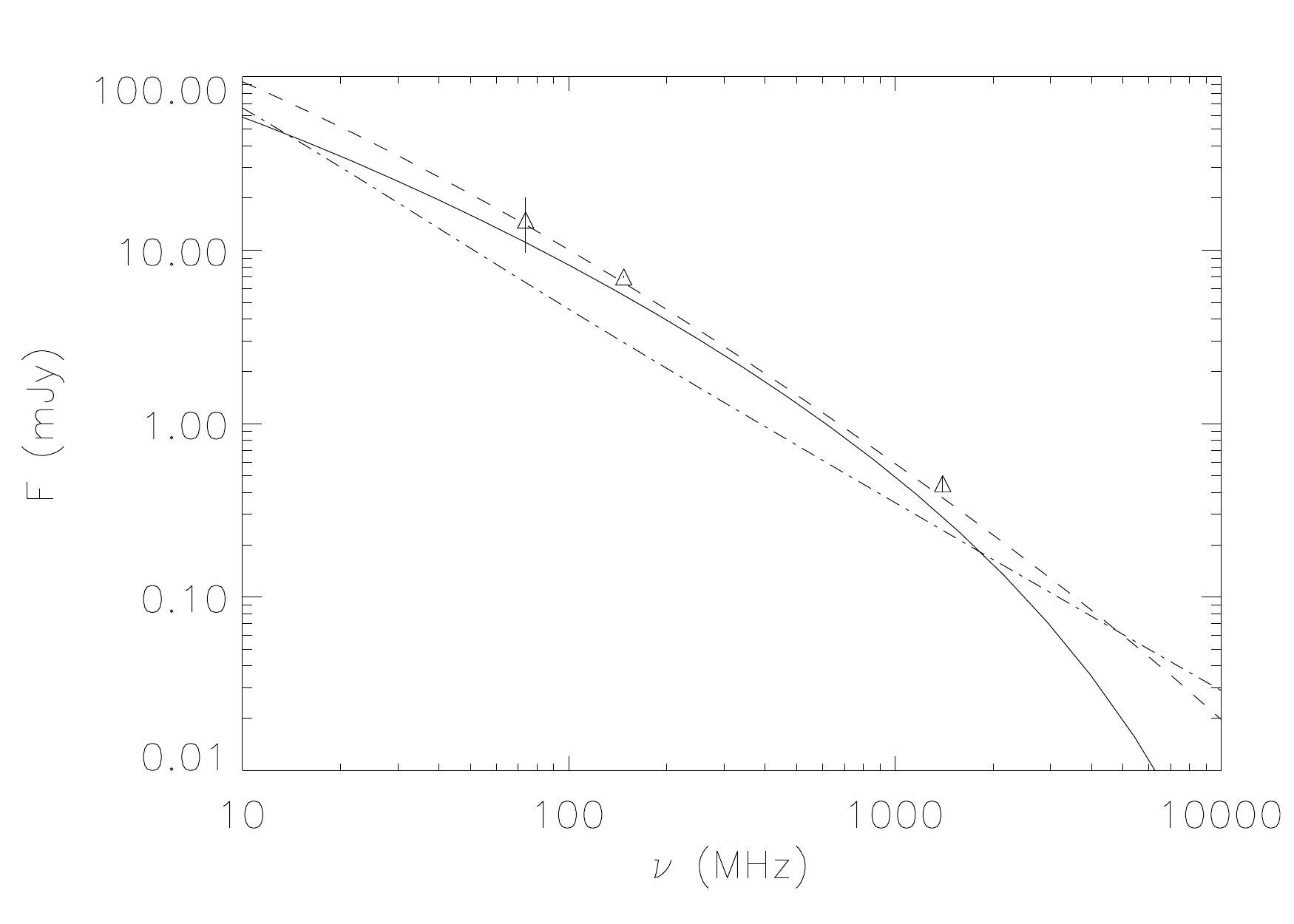}
}
\end{tabular}
\caption{Spectrum of the radio halo flux in the region P2 of A520 for a magnetic field of 3.8 $\mu$G (upper panel) and 5 $\mu$G (lower panel) and for: \textit{i)} a neutralino model with mass $M_\chi=9$ GeV, annihilation final state $\tau^+\tau^-$, ${\cal B} \times \langle \sigma v \rangle=6\times10^{-25}$ cm$^3$ s$^{-1}$ (solid line); \textit{ii)} a neutralino model with mass $M_\chi=43$ GeV, annihilation final state $b \bar b$, ${\cal B} \times \langle \sigma v \rangle=4\times10^{-24}$ cm$^3$ s$^{-1}$ (dashed line); \textit{iii)} a hadronic model with proton spectral index $s_p=2.2$ and a pressure ratio $P_{CR}/P_{th}=4\%$ (dot-dashed line). Data are from Table \ref{p2fluxes}.}
\label{halo.dm+sec_p2}
\end{figure}

In Fig.\ref{halo.dm+sec_p2} we show the radio emission produced in this region for the hadronic and DM models, using for the hadronic model a reference pressure ratio $P_{CR}/P_{th}=4\%$, higher than the one providing the best fitting to the whole radio halo. Even with this value, the DM emission appears to be dominant with respect to the hadronic one. This is because in a small region the contribution of the central peak of the DM emission is more important than the one provided from the flat profile of non-thermal protons. The expected DM radio emission is smaller than the flux estimated from the surveys in this region by a factor of 1.3--1.4.

We also found that, by using instead a magnetic field of the order of 5 $\mu$G, it is possible to obtain a radio flux produced by DM similar to the observed one (lower panel of Fig.\ref{halo.dm+sec_p2}). Since we don't have information on the magnetic field in this cluster, this possibility can not be excluded. We note also that the model with neutralino mass 43 GeV reproduces the spectral shape of the radio emission better than the 9 GeV model.

Therefore, the DM emission in the P2 region can be dominant on the hadronic one if the normalization of the DM emission is the same than the one that can reproduce the radio halo flux in Coma, and this suggests that the contribution of DM to the radio emission in the cluster, if present, can be found in this region.

\section{High energy emission}

In this Section we discuss if the emission produced in the high energy bands (X-ray and gamma ray) is suitable to be detected with present or forthcoming instruments, in order to understand if with this kind of observations it is possible to discriminate between the different models and derive more information on the properties and the origin of cosmic ray particles in A520 and in the P2 region.

\begin{figure}
\centering
\begin{tabular}{c}
\vbox{
\includegraphics[width=\columnwidth]{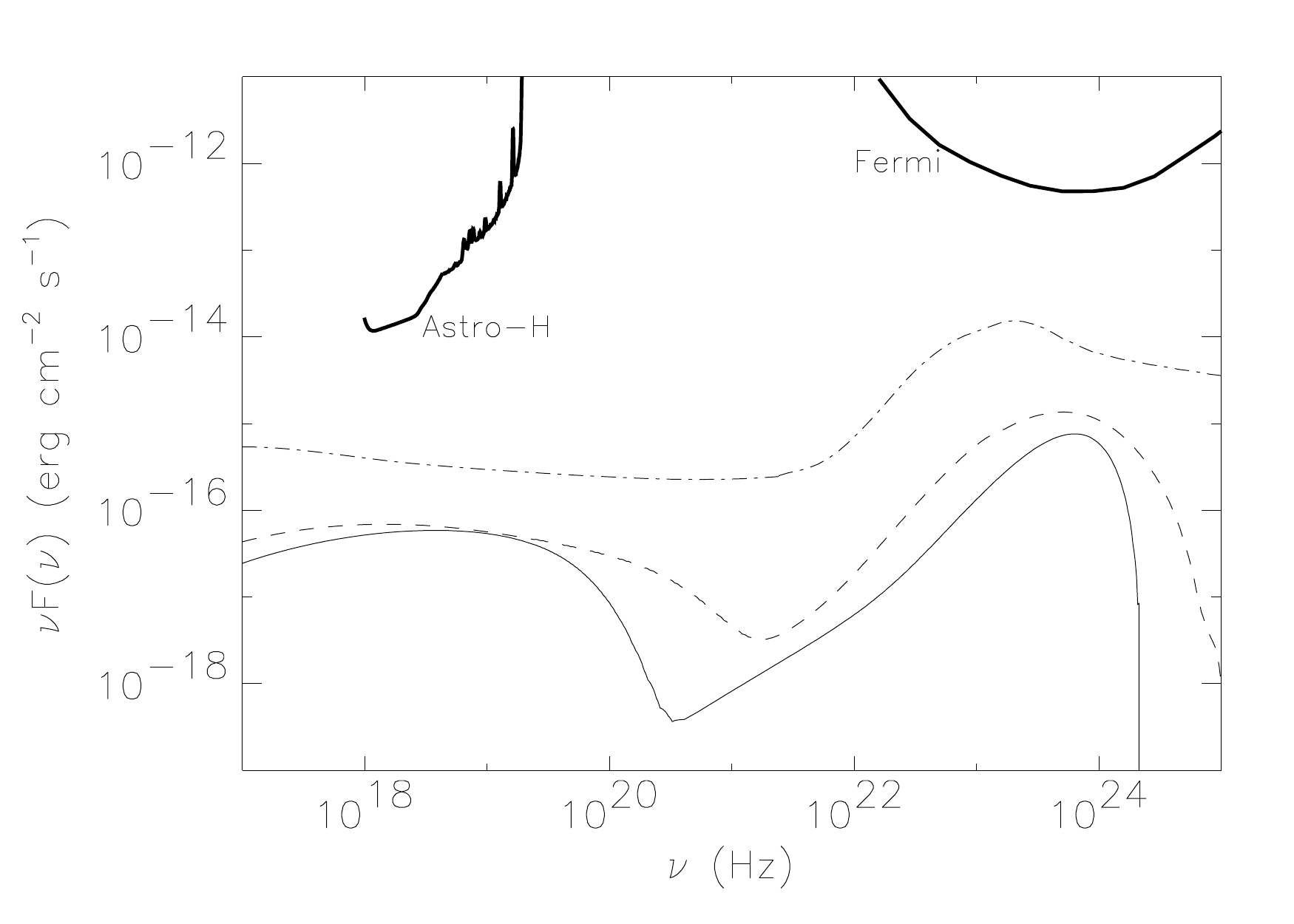}
\includegraphics[width=\columnwidth]{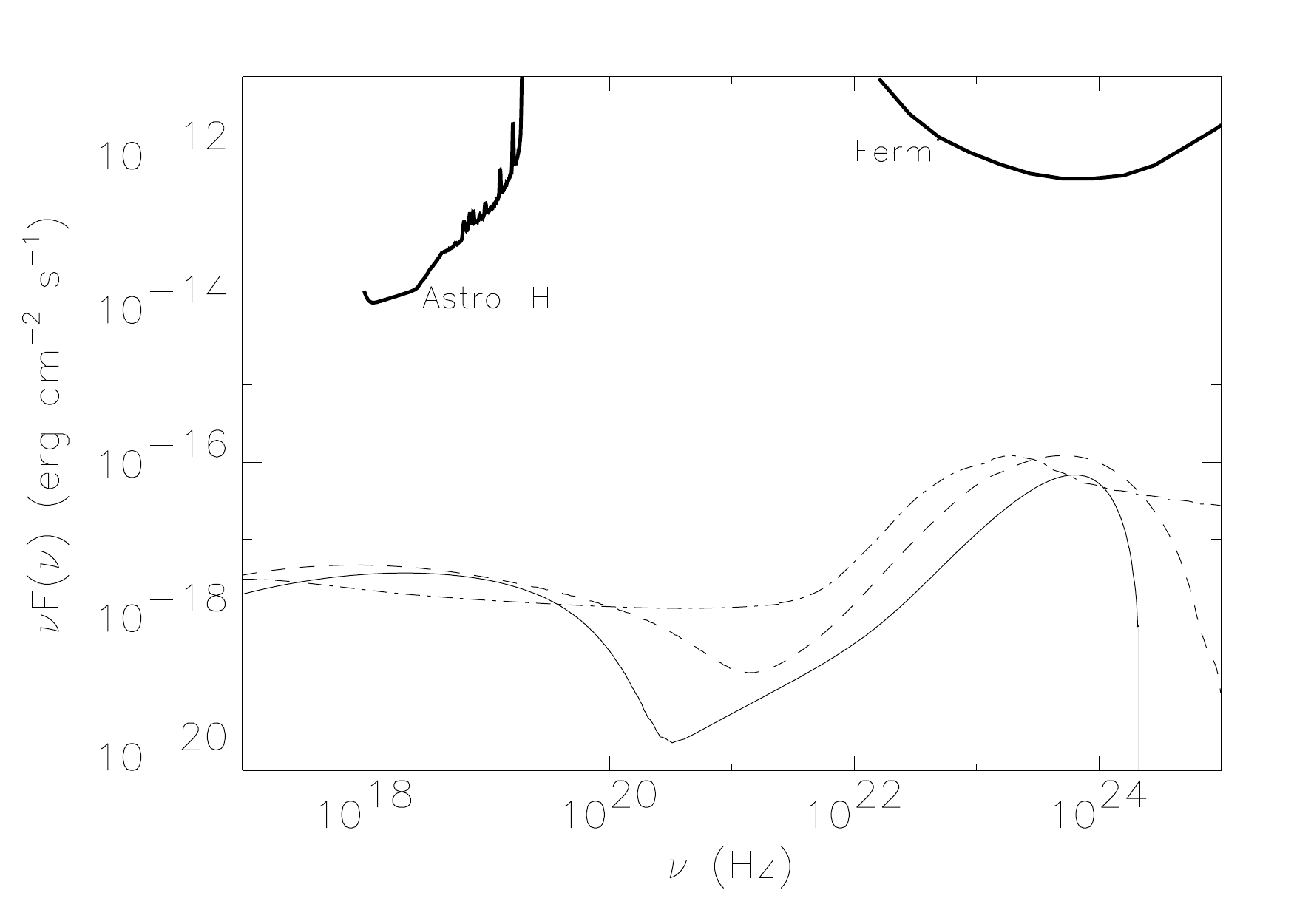}
}
\end{tabular}
\caption{Spectral energy distribution of the high energy emission from the whole cluster A520 (upper panel) and the P2 region (lower panel) for: \textit{i)} a neutralino model with mass $M_\chi=9$ GeV, annihilation final state $\tau^+\tau^-$, ${\cal B} \times \langle \sigma v \rangle=6\times10^{-25}$ cm$^3$ s$^{-1}$ (solid line); \textit{ii)} a neutralino model with mass $M_\chi=43$ GeV, annihilation final state $b \bar b$, ${\cal B} \times \langle \sigma v \rangle=4\times10^{-24}$ cm$^3$ s$^{-1}$ (dashed line); \textit{iii)} a hadronic model with proton spectral index $s_p=2.2$ and pressure ratios $P_{CR}/P_{th}=2.6\%$ in the whole cluster and $4\%$ in the P2 region (dot-dashed line).
We also show the sensitivity curves of Astro-H for 100 ks of time integration (from http://astro-h.isas.jaxa.jp/researchers/sim/sensitivity.html) and Fermi-LAT for 10 yrs (from Funk \& Hinton 2013).}
\label{highe.dm+sec}
\end{figure}

In Fig.\ref{highe.dm+sec} we show the high-energy emission given by the sum of the Inverse Compton Scattering of the CMB photons and non-thermal bremsstrahlung with the thermal ions from the secondary electrons, and the gamma ray emission produced by neutral pions decay, for the three models we used in the past Section, and for the two regions we considered: the whole cluster (upper panel), and the P2 region (lower panel). 

We can see that in the whole cluster the hadronic emission dominates over the emission of DM origin along the whole spectrum, whereas in the P2 region the DM emission is stronger than the hadronic one in the X-ray band (between 0.3 and 400 keV for the 43 GeV model, and between 1 and 150 keV for the 9 GeV model), while in the gamma rays the hadronic emission is generally stronger than the DM one, except for the spectral region between 1 and 10 GeV for the 43 GeV mass model, and between 2 and 4 GeV for the 9 GeV model. However, all these emissions are well below the sensitivity of present and forthcoming instruments (in the same Figure we show, for reference, the sensitivities of Astro-H for 100 ks of integration and Fermi-LAT for 10 yrs) even for the whole cluster, so we don't expect that observations in these bands can help to discriminate the origin of the non-thermal emission in A520.

\section{Discussion}

Like in the Bullet cluster (Marchegiani \& Colafrancesco 2015), in A520 the emission of DM origin probably is not producing the whole radio halo, but can be dominant in some regions of the cluster. In the whole cluster the baryonic emission is expected to be dominant on the DM one. 

In the SW part of the cluster a possible source of acceleration of cosmic rays is the bow shock associated to the BCG at the edge of the radio halo that is moving outwards; however, Wang et al. (2018), using Chandra X-ray observations, found that the Mach number associated to this shock is $\mach = 2.4^{+0.4}_{-0.2}$; according to the Diffusive Shock Acceleration theory (Drury 1983), particles accelerated by a shock with this Mach number should produce a radio emission with a spectral index equal to or steeper than 1.3, that is steeper than the values of 1.1-1.2 found in this region (Vacca et al. 2014). Another possible source of cosmic rays is the BCG itself, that can have injected the particles during its passage through the Intra Cluster Medium (ICM). Since the particles are diffused up a distance of the order of 500 kpc (much higher than the typical diffusion length of electrons in magnetized plasma in galaxy clusters, that is of the order of 50 kpc, see, e.g., Blasi \& Colafrancesco 1999), this emission probably is of hadronic origin.

In the NE part of the cluster we have identified a peak of the radio halo that is located very close to one of the peaks of the DM distribution (the P2 peak in Clowe et al. 2012). We have found that in this region the emission of DM origin is expected to dominate on the emission of baryonic origin, if the DM has properties similar to the ones providing the best fitting to the spectrum of the radio halo in Coma (Marchegiani \& Colafrancesco 2016), when optimistic, but possible, values for the annihilation cross section and the substructure boosting factor are assumed. Therefore, if a contribution from DM to the radio halo is present in this cluster, it needs to be searched in this region. We note that, while in the Coma cluster the DM model providing the best fitting to the radio halo spectrum is the one with mass 9 GeV and annihilation final state $\tau^+\tau^-$, in the P2 region of A520 the model providing the best fitting to the radio spectrum is the one with mass 43 GeV and annihilation final state $b \bar b$. Therefore other studies, even considering similar situations in other clusters, need to be performed in order to clarify the strenght of this result.

The whole NE part of the radio halo is more extended than the P2 region alone, that has a radius similar to the core radius expected for a DM halo having the corresponding mass ($\sim115$ kpc); therefore other mechanisms that can accelerate the electrons on larger scales need to be present. The apparent separation of the NE part of the radio halo from the SW part suggests that other mechanisms different from the injection from the BCG can take place. As suggested by Wang et al. (2018), a possible mechanism is the reacceleration caused by merger-induced turbulences, as suggested also by the high values of the temperature found in this region (Govoni et al. 2004; Wang et al. 2016). Another possibility is that an important role is played by the radio galaxy labelled with D in Vacca et al. (2014): in the temperature map of Govoni et al. (2004) it is possible to see that a region of hot gas is present on the west of the galaxy (i.e. in the probable direction of its motion), and is bending north and south of the galaxy. It is therefore possible that the bow shock produced by the galaxy motion is providing energy to the cosmic ray electrons by diffusive shock acceleration or adiabatic compression. A shock is not visible in this region of the X-ray maps (Wang et al. 2016), but this fact can be explained because this radio galaxy is probably part of a subcluster that is falling on the main cluster along a direction close to the line of sight (Proust et al. 2000). It is also possible that the radio galaxy, by crossing a merger shock front, is giving origin to a relic-like structure (like in the case of Perseus cluster, see Pfrommer \& Jones 2011) that would appear roundish because associated to a shock front produced in a merger happening close to the line of sight, as suggested by Vacca et al. (2014).

\section{Conclusions}

In this paper we have found that in the cluster A520 the radio emission produced by DM is expected to be lower than the emission of baryonic origin. 
However, DM can have an important role in the region around the subhalo P2, where a peak of the radio emission is detected.

We have estimated the radio flux in the P2 region, and found that DM models with the same properties as the ones providing a good fitting to the radio halo in the Coma cluster (Marchegiani \& Colafrancesco 2016) predict a radio emission similar to the observed one, for a magnetic field of the order of 5 $\mu$G.
The best fitting model suggested by the spectral shape of the radio emission in the P2 region is for a neutralino with mass 43 GeV and annihilation final state $b \bar b$; this situation is different from the Coma case, where the favourite model was for mass 9 GeV and final state $\tau^+ \tau^-$.

In the SW part of the cluster, a possibile origin of the radio halo is given by hadronic interactions, where the cosmic rays are injected by the BCG moving through the ICM.
In the NE part, other than the DM in the P2 region, an important role can be played by the radio galaxy D that is probably interacting with the ICM, emitting cosmic rays and/or producing a bow shock.

Finally, the expected high energy emission from the whole cluster and the P2 region is well below the sensitivities of present and forthcoming instruments, and therefore measures in high energy spectral bands do not appear to be suitable to obtain better information on the non-thermal emission in A520.

%%%%%%%%%%%%%%%%%%%%%%%%%%%%%%%%%%%%%%%%%%%%%%%%%%%%

\section*{Acknowledgments}
SC passed away during the final stages of the preparation of this paper; he gave a valuable contribution to its realization and we strongly thank him. We also thank Prof. J. Carter for his support in a difficult moment.
This work is based on the research supported by the South African Research Chairs Initiative
of the Department of Science and Technology and National Research Foundation of South
Africa (Grant No 77948). PM and NK acknowledge support from the Department of Science and Technology/National Research Foundation
(DST/NRF) Square Kilometre Array (SKA) post-graduate bursary initiative under the same Grant.

%%%%%%%%%%%%%%%%%%%%%%%%%%%%%%%%%%%%%%%%%%%%%%%%%

%%%%%%%%%%%%%%%%%%%%%%%%%%%%%%%%%%%%%%%%%%%%%%%%%%%%%%%%

\bsp

\label{lastpage}


\begin{thebibliography}{99}

\bibitem[\protect\citeauthoryear{}{}]{Abaza2016}
Abazajian K.N., Keeley R.E., 2016, Phys. Rev. D, 93, 083514

\bibitem[\protect\citeauthoryear{}{}]{Ade2016}
Ade P.A.R. et al., 2016, A\&A, 594, A13

\bibitem[\protect\citeauthoryear{}{}]{BlasiCola1999} 
Blasi P., Colafrancesco S., 1999, Astroparticle Physics, 12, 169

\bibitem[\protect\citeauthoryear{}{}]{Bonafede2010}
Bonafede A., Feretti L., Murgia M., Govoni F., Giovannini G., Dallacasa D., Dolag K., Taylor G.B., 2010, A\&A, 513, A30

\bibitem[\protect\citeauthoryear{}{}]{Brown2011}
Brown S., Rudnick L., 2011, MNRAS, 412, 2

\bibitem[\protect\citeauthoryear{}{}]{Bullock2001}
Bullock J.S. et al., 2001, MNRAS, 321, 559

\bibitem[\protect\citeauthoryear{}{}]{CavaFusco76} 
Cavaliere A., Fusco-Femiano R., 1976, A\&A, 49, 137

\bibitem[\protect\citeauthoryear{}{}]{Clowe2006}
Clowe D., Bradac M., Gonzalez A.H., Markevitch M., Randall S.W., Jones C., Zaritsky D., 2006, ApJ, 648, L109

\bibitem[\protect\citeauthoryear{}{}]{Clowe2012}
Clowe D., Markevitch M., Bradac M., Gonzalez A.H., Chung S.M., Massey R., Zaritsky D., 2012, ApJ, 758, 128

\bibitem[\protect\citeauthoryear{}{}]{Colafrancesco2006}
Colafrancesco S., Profumo S., Ullio P., 2006, A\&A, 455, 21

\bibitem[\protect\citeauthoryear{}{}]{Colafrancesco2015}
Colafrancesco S., Marchegiani P., Beck G., 2015, JCAP, 02, 032

\bibitem[\protect\citeauthoryear{}{}]{Condon1998}
Condon J.J., Cotton W.D., Greisen E.W., Yin Q.F., Perley R.A., Taylor G.B., Broderick J.J., 1998, AJ, 115, 1693

\bibitem[\protect\citeauthoryear{}{}]{Drury1983} 
Drury L.O., 1983, Reports on Progress in Physics, 46, 973 

\bibitem[\protect\citeauthoryear{}{}]{Feretti2004}
Feretti L., Orr{\`u} E., Brunetti G., Giovannini G., Kassim N., Setti G., 2004, A\&A, 423, 111

\bibitem[\protect\citeauthoryear{}{}]{Feretti2012}
Feretti L., Giovannini G., Govoni F., Murgia M., 2012, A\&ARv, 20, 54

\bibitem[\protect\citeauthoryear{}{}]{Funk2013}
Funk S., Hinton J.A., 2013, Astroparticle Physics, 43, 348

\bibitem[\protect\citeauthoryear{}{}]{Giovannini1993}
Giovannini G., Feretti L., Venturi T., Kim K.-T., Kronberg P.P., ApJ, 406, 399

\bibitem[\protect\citeauthoryear{}{}]{Gondolo2004}
Gondolo P. et al., 2004, Journal of Cosmology and Astroparticle Physics, 07, 008

\bibitem[\protect\citeauthoryear{}{}]{Govoni2001}
Govoni F., Feretti L., Giovannini G., B{\"o}hringer H., Reiprich T.H., Murgia M., 2001, A\&A, 376, 803

\bibitem[\protect\citeauthoryear{}{}]{Govoni2004}
Govoni F., Markevitch M., Vikhlinin A., van Speybroeck L., Feretti L., Giovannini G., 2004, ApJ, 605, 695

\bibitem[\protect\citeauthoryear{}{}]{Huber2013}	
Huber B., Tchernin C., Eckert D., Farnier C., Manalaysay A., Straumann U., Walter R., 2013, A\&A, 560, A64

\bibitem[\protect\citeauthoryear{}{}]{Intema2017}		
Intema H.T., Jagannathan P., Mooley K.P., Frail D.A., 2017, A\&A, 598, A78

\bibitem[\protect\citeauthoryear{}{}]{Lane2014}
Lane W.M., Cotton W.D., van Velzen S., Clarke T.E., Kassim N.E., Helmboldt J.F., Lazio T.J.W., Cohen A.S., 2014, MNRAS, 440, 327

\bibitem[\protect\citeauthoryear{}{}]{Mahdavi2007}
Mahdavi A., Hoekstra H., Babul A., Balam D.D., Capak P.L., 2007, ApJ, 668, 806

\bibitem[\protect\citeauthoryear{}{}]{Marchegiani2015}
Marchegiani P., Colafrancesco S., 2015, MNRAS, 452, 1328

\bibitem[\protect\citeauthoryear{}{}]{Marchegiani2016}
Marchegiani P., Colafrancesco S., 2016, JCAP, 11, 033

\bibitem[\protect\citeauthoryear{}{}]{Marchegiani2007}
Marchegiani P., Perola G.C., Colafrancesco S., 2007, A\&A, 465, 41

\bibitem[\protect\citeauthoryear{}{}]{Markevitch2005}
Markevitch M., Govoni F., Brunetti G., Jerius D., 2005, ApJ, 627, 733

\bibitem[\protect\citeauthoryear{}{}]{MoskaStrong98} 
Moskalenko I.V., Strong A.W., 1998, ApJ, 493, 694

\bibitem[\protect\citeauthoryear{}{}]{nfw1996}
Navarro J.F., Frenk C.S., White S.D.M., 1996, ApJ, 462, 563

\bibitem[\protect\citeauthoryear{}{}]{Pfrommer2011} 
Pfrommer C., Jones T.W., 2011, ApJ, 730, 22

\bibitem[\protect\citeauthoryear{}{}]{Prokhorov2014} 
Prokhorov D.A., Churazov E.M., 2014, A\&A, 567, A93

\bibitem[\protect\citeauthoryear{}{}]{Proust2000}
Proust D., Cuevas H., Capelato H.V., Sodr{\'e} L.Jr., Tom{\'e} Lehodey B., Le F{\`e}vre O., Mazure A., 2000, A\&A, 355, 443

\bibitem[\protect\citeauthoryear{}{}]{Roettiger1997}
Roettiger K., Loken C., Burns J.O., 1997, ApJS, 109, 307

\bibitem[\protect\citeauthoryear{}{}]{Sarazin1999}
Sarazin C.L., 1999, ApJ, 520, 529

\bibitem[\protect\citeauthoryear{}{}]{Vacca2014}
Vacca V., Feretti L., Giovannini G., Govoni F., Murgia M., Perley R.A., Clarke T.E., 2014, A\&A, 561, A52 

\bibitem[\protect\citeauthoryear{}{}]{Wang2016}
Wang Q.H.S., Markevitch M., Giacintucci S., 2016, ApJ, 833, 99

\bibitem[\protect\citeauthoryear{}{}]{Wang2018}
Wang Q.H.S., Giacintucci S., Markevitch M., 2018, ApJ, 856, 162


\end{thebibliography}
\end{document}